# Monocyte and T-lymphocyte trans-endothelial migration in relation to cardiovascular disease: some alternative boundary conditions in a model recently proposed by Little *et al.* (*PLoS Comput Biol* 2009 5(10) e1000539)

## Mark P. Little[a, b]

[a]Department of Epidemiology and Biostatistics, School of Public Health, Imperial College Faculty of Medicine, Norfolk Place, London W2 1PG, UK
[b]To whom all correspondence should be addressed at: tel +44 (0)20 7594 3312; Fax +44 (0)20 7402 2150;    Email mark.little@imperial.ac.uk

**Abstract**
We consider a slight modification to the monocyte and T-lymphocyte boundary conditions of Little *et al.* (1) and derive alternative parameter estimates. No changes to the results and conclusions of the paper of Little *et al.* (1) are implied.

**Introduction and Results**
We shall consider a slight modification of the boundary monocyte and T-lymphocyte flux given by equations (17) and (18) in Little *et al.* (1) namely:

$$J_m(C, m_L, m_I) = \begin{cases} D_m \beta_m (C, m_L, m_I) = D_m \begin{cases} \beta_{m,0}[m_L - m_I] + \\ \beta_{m,1}[m_L - m_I + \kappa \cdot m_L] \mathrm{Cl}_{C > C_{Tm}} \end{cases} \text{ on } \Gamma_1 \\ 0 \text{ otherwise} \end{cases} \quad (1)$$

$$J_T(C, T_L, T_I) = \begin{cases} D_T \beta_T (C, T_L, T_I) = D_T \begin{cases} \beta_{T,0}[T_L - T_I] + \\ \beta_{T,1}[T_L - T_I + \kappa T_L] \mathrm{Cl}_{C > C_{TT}} \end{cases} \text{ on } \Gamma_1 \\ 0 \text{ otherwise} \end{cases} \quad (2)$$

where $m_L, m_I$ are the monocyte concentrations in the lumen and intima, respectively, and $T_L, T_I$ are the T-lymphocyte concentrations in the lumen and intima, respectively. The fundamentally linear form of these is inspired by data in Takaku *et al.* (2) and Klouche *et al.* (3). If we set $m_T = m_L + m_I$, then in the context of the experimental system studied by Takaku *et al.* (2), with a fixed monocyte population, a proportion of which transmigrates, this leads to the differential equation:

$$\frac{\partial m_L}{\partial t} = -D_m \left[ \beta_{m,0}[2m_L - m_T] + \beta_{m,1}[(2+\kappa)m_L - m_T] \mathrm{Cl}_{C > C_{Tm}} \right] \quad (3)$$

Setting $P = m_L / m_T$ we have that:

$$\frac{\partial P}{\partial t} = -D_m \left[ \beta_{m,0}[2P - 1] + \beta_{m,1}[(2+\kappa)P - 1] \mathrm{Cl}_{C > C_{Tm}} \right] \quad (4)$$

With the initial condition $P(0) = 1$ this has solution:

$$P(t) = \exp\left[-2D_m t\left(\beta_{m,0} + \beta_{m,1}C\left[\frac{\kappa}{2}+1\right]1_{C>C_{Tm}}\right)\right] + 0.5\frac{D_m\left(\beta_{m,0} + \beta_{m,1}C1_{C>C_{Tm}}\right)}{D_m\left(\beta_{m,0} + \beta_{m,1}C\left[\frac{\kappa}{2}+1\right]1_{C>C_{Tm}}\right)}$$

$$= \exp\left[-2D_m t\left(\beta_{m,0} + \beta_{m,1}C\left[\frac{\kappa}{2}+1\right]1_{C>C_{Tm}}\right)\right] + 0.5\frac{\beta_{m,0} + \beta_{m,1}C1_{C>C_{Tm}}}{\beta_{m,0} + \beta_{m,1}C\left[\frac{\kappa}{2}+1\right]1_{C>C_{Tm}}}$$

(5)

We fit this by least squares to the data in Figure 4 of Takaku *et al.* (2) (the inferred values on proportion of non-migrant monocytes are given in Table 1 below) to obtain values of $D_m\beta_{m,0}, D_m\beta_{m,1}, \kappa$. We divide the first two of these by $D_m = 3.03 \times 10^{-15}\,\mathrm{m^2 s^{-1}}$ (as derived by Little *et al.* (1)) to give $\beta_{m,0} = 1.54 \times 10^9\,\mathrm{m^{-2}}$, $\beta_{m,1} = 5.37 \times 10^{23}\,\mathrm{M^{-1}ml\,m^{-2}}$ and $\kappa = 0.787$. Both $\beta_{m,0}, \beta_{m,1}$ values are somewhat larger than those originally derived by Little *et al.* (1), namely $\beta_{m,0} = 1.72 \times 10^8\,\mathrm{m^{-2}}$ and $\beta_{m,1} = 2.53 \times 10^{19}\,\mathrm{M^{-1}ml\,m^{-2}}$. [Note: there was a typographical error in the paper where this value was recorded as $\beta_{m,1} = 2.5 \times 10^{13}\,\mathrm{M^{-1}ml\,m^{-2}}$.]

As in the paper of Little *et al.* (1) we scale these values to corresponding ones for T-lymphocytes using the relations $D_T\beta_{T,0} = D_m\beta_{m,0} \cdot \frac{10.4}{11.9} = 4.08 \times 10^{-6}\,\mathrm{s^{-1}}$, $D_T\beta_{T,1} = D_m\beta_{m,1} \cdot \frac{(35.8-10.4)}{(47.2-11.9)} = 1.17 \times 10^9\,\mathrm{M^{-1}ml\,s^{-1}}$, using values given in Figure 4 of Klouche *et al.* (3) (and reproduced in Table 2 below). Finally dividing these by $D_T = 6.21 \times 10^{-15}\,\mathrm{m^2 s^{-1}}$ (as derived by Little *et al.* (1)) we obtain $\beta_{T,0} = 6.58 \times 10^8\,\mathrm{m^{-2}}$, $\beta_{T,1} = 1.88 \times 10^{23}\,\mathrm{M^{-1}ml\,m^{-2}}$ and (as for monocytes) $\kappa = 0.787$. Both $\beta_{T,0}, \beta_{T,1}$ values are somewhat larger than those originally derived by Little *et al.* (1), namely $\beta_{T,0} = 7.33 \times 10^7\,\mathrm{m^{-2}}$ and $\beta_{T,1} = 8.85 \times 10^{18}\,\mathrm{M^{-1}ml\,m^{-2}}$. [Note: again, there was a typographical error in the paper where this value was recorded as $\beta_{T,1} = 8.9 \times 10^{12}\,\mathrm{M^{-1}ml\,m^{-2}}$.] Some of the differences may relate to the fact that the original data-fitting was to absolute numbers rather than (as here) to the ratios, $P = m_L / m_T$, which we do because of evidence that the total number of cells does not appear to be constant.

As the results of the paper of Little *et al.* (1) assumed zero boundary flux of monocytes and T-lymphocytes, these possible changes in the form of the model, and to the derived coefficients, have no implications for the results and conclusions of that paper.

**Table 1. Migrant and non-migrant monocyte values derived from Figure 4 of Takaku *et al.* (2)**

| MCP 1 concentration (nM ml$^{-1}$) | Days after start | Migrant cells | Non-migrant cells | Migrant proportion (%) | Non-migrant proportion (%) |
|---|---|---|---|---|---|
| 0.0 | 1 | 7300 | 63700 | 10.3 | 89.7 |
| 0.0 | 4 | 27800 | 55200 | 33.5 | 66.5 |
| 0.0 | 7 | 19000 | 27400 | 40.9 | 59.1 |
| 0.03 | 1 | 61800 | 38800 | 61.5 | 38.5 |
| 0.03 | 4 | 54100 | 34800 | 60.9 | 39.1 |
| 0.03 | 7 | 66600 | 28500 | 70.0 | 30.0 |

**Table 2. Migrant monocyte and T-lymphocyte values derived from Figure 4 of Klouche *et al.* (3)**

| Group | Migrant monocyte proportion (%) | Migrant T-lymphocyte proportion (%) |
|---|---|---|
| Control | 11.9 | 10.4 |
| 25 µg ml$^{-1}$ E-LDL | 47.2 | 35.8 |